\documentclass[twocolumn,aps,prl,amssymb,superscriptaddress]{revtex4-1}
\usepackage{amsmath}
\usepackage{graphicx}
\DeclareGraphicsExtensions{.pdf,.eps,.png,.jpg,.mps}
\usepackage{epstopdf}
\usepackage{threeparttable}
\usepackage{multirow}
\usepackage{color}
\begin{document}
\title{Graphene Phase Modulator}
\author{V. Sorianello$^1$, M. Midrio$^2$,G. Contestabile$^3$,I. Asselberg$^4$,J. Van Campenhout$^4$,C. Huyghebaerts$^4$,I. Goykhman$^5$,A. K. Ott$^5$,A. C. Ferrari$^5$,M.Romagnoli}
\affiliation{Consorzio Nazionale per le Telecomunicazioni (CNIT), National Laboratory of Photonic Networks, Via G. Moruzzi 1, 56124 Pisa, Italy
\\$^2$Consorzio Nazionale per le Telecomunicazioni (CNIT), University of Udine, Via delle Scienze 206, 33100 Udine, Italy
\\$^3$Consorzio Nazionale per le Telecomunicazioni (CNIT), Scuola Superiore Sant'Anna, via G. Moruzzi 1, 56124 Pisa, Italy
\\$^4$IMEC, Kapeldreef 75, 3001 Leuven, Belgium
\\$^5$Cambridge Graphene Centre, Cambridge University, 9 JJ Thomson Avenue, Cambridge CB3 OFA, UK}
\begin{abstract}
We demonstrate a 10Gb/s Graphene Phase Modulator (GPM) integrated in a Mach-Zehnder interferometer configuration. This is a compact device, with a phase-shifter length of only 300$\mu$m, and 35dB extinction ratio. The GPM has modulation efficiency of 0.28Vcm, one order of magnitude higher compared to state-of-the-art depletion p-n junction Si phase modulators. Our GPM operates with 2V peak-to-peak driving voltage in a push-pull configuration, and it has been tested in a binary transmission of a non-return-to-zero data stream over 50km single mode fibre. This device is the key building block for graphene-based integrated photonics, enabling compact and energy efficient hybrid Si-graphene modulators for telecom, datacom and other applications.
\end{abstract}
\maketitle
Optical communication systems for telecom and datacom implement high-order digital modulation schemes, such as amplitude or phase-shift keying, in order to enhance spectral efficiency and increase the data transmission capacity of telecommunication networks\cite{Seimetz2009}. This is possible by encoding several bits of information in fewer symbols, leading to the same amount of data, but transmitted at lower speed, resulting in reduced spectral bandwidth. Spectral narrowing is a key technique for optimizing data capacity in modern wavelength-division multiplexing (WDM)\cite{Nakazawa2010} allowing higher data rates, improved spectral efficiency and robust tolerances to transmission impairments, such as chromatic dispersion\cite{Seimetz2009}.

The modulator is a key component in optical communications\cite{Agrawal2010}. To encode the information in complex modulation formats, phase modulation (PM) is needed\cite{Seimetz2009}. In integrated Si photonics, this is achieved by exploiting the free-carriers plasma dispersion effect\cite{Soref1987}, whereby changes in electron and hole densities result in changes in the Si refractive index and absorption. In state-of-the-art Si modulators\cite{Xiao2013,Streshinsky2013, Denoyer2015,Reed2013,Dong2012,Fresi2016,Milivojevic2013} the complex index change can be driven by p-n junctions\cite{Xiao2013,Streshinsky2013,Denoyer2015,Reed2013,Dong2012,Fresi2016} or capacitors\cite{Milivojevic2013} built in Si waveguides, and applying a voltage to deplete or accumulate free-carriers interacting with the propagating light. High-speed for simple and complex modulation formats has been demonstrated\cite{Xiao2013,Streshinsky2013,Denoyer2015,Reed2013,Dong2012,Fresi2016,Milivojevic2013} either with a depleted p-n junction under reverse bias\cite{Xiao2013,Streshinsky2013,Denoyer2015,Reed2013,Dong2012,Fresi2016}, or with a Si-insulator-Si (SIS) capacitor\cite{Milivojevic2013} operating in the accumulation regime. However, the Si free-carrier effect in the depleted p-n junction\cite{Soref1987} typically requires mm-size devices for accumulating a $\pi$-phase shift\cite{Reed2013} with a 1-3V driving voltage. This limits the modulation efficiency, defined as the product of the $\pi$-phase-shift voltage and length (V$_{\pi}$L)\cite{Reed2008}, and increase energy consumption (energy per bit)\cite{Miller2012}. Another challenge is that plasma dispersion is always accompanied by absorption\cite{Soref1987}. As a result, there is a tradeoff between optical loss, V$_{\pi}$L and footprint, since higher free carriers densities with stronger plasma dispersion come at the expense of increased optical absorption\cite{Reed2008} or limited dynamic range, i.e. the maximum intensity or phase change for a given driving voltage\cite{Reed2008}. Therefore, novel solutions with increased V$_{\pi}$L, reduced optical loss, and miniaturized footprints are needed.

Graphene is appealing for photonics and optoelectronics because it offers a wide range of advantages compared to other materials, such as Si and other semiconductors\cite{Bonaccorso2010,Kim2011,Ferrari2015,CastroNeto}. In particular, it is ideally suited for integration into Si photonics\cite{Bonaccorso2010, Hanson2008,Mak2008} due to its large optical modulation\cite{Liu2011,Liu2012,Hu2016,Phare2015,Sorianello2015}, broadband photodetection\cite{Koppens2014,Goykhman2016}, high-speed operation\cite{Phare2015,Koppens2014}, and CMOS (complementary metal oxide semiconductor) compatibility\cite{Liu2011,Liu2012,Hu2016,Phare2015,Sorianello2015,Koppens2014,Goykhman2016}. To date, several graphene based amplitude modulators have been reported based on electro-absorption\cite{Liu2011,Liu2012,Hu2016,Phare2015}. However, graphene phase modulators (GPMs) are necessary for all functionalities requiring phase change, such as binary and complex modulation formats, switching, phased arrays, etc. GPMs are also expected to have an improved efficiency-loss figure of merit compared to Si based modulators\cite{Sorianello2015}. Graphene integration with Si photonics opens a new paradigm for developing compact, efficient and low-loss integrated PMs outperforming state-of-the-art Si devices\cite{Xiao2013, Streshinsky2013,Denoyer2015,Reed2013, Dong2012,Fresi2016,Milivojevic2013}. Due to its unique opto-electronic properties, optical losses in single layer graphene (SLG) can be electrically suppressed due to Pauli blocking\cite{Wang2008, Falkovski2008, Stauber2008}. This modulates the absorption, with a refractive index change much larger than that the free carriers effect in Si\cite{Sorianello2015}. SLG on Si waveguides with effective refractive index (i.e. the ratio between the light phase velocity in the waveguide and in vacuum) changes larger than 10$^{-3}$ have been demonstrated\cite{Sorianello2016}, about ten-times larger than state-of-the-art Si p-n junction waveguides in depletion mode\cite{Reed2008}. Consequently, SLG can provide a unique combination of strong electro-refractive effect and optical transparency when operated in the Pauli blocking regime.

\begin{figure}
\centerline{\includegraphics[width=80mm]{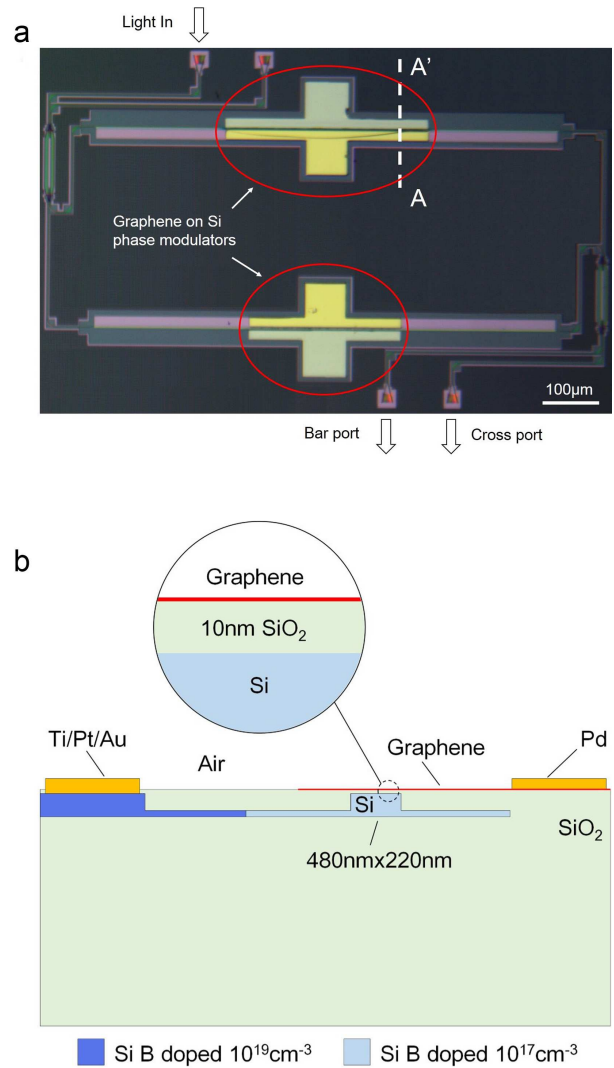}}
\caption{a) Optical micrograph of the MZI modulator. The top arm has a 400$\mu$m SLG on a Si GPM, the bottom arm has a 300$\mu$m SLG on a Si GPM. b)Cross-section of the GPM in the section A-A' of (a). The ridge waveguide has a 480x220nm core with a 60nm thick lateral slab. The waveguide is B doped with a concentration$\sim$10$^{17}$cm$^{-3}$ (pale blue) in the core and in the slab up to 1$\mu$m from the core. Heavy B doping$\sim$10$^{19}$cm$^{-3}$ (violet) is present in the contact region in order to improve the contact resistance on Si. The waveguide is embedded in a SiO$_2$ cladding (light green) 2$\mu$m thick at the bottom and 10nm thick on the top of the waveguide. The SLG (red) is placed on top of the waveguide and extends$\sim$0.5$\mu$m beyond the waveguide core on the left side. Different metallizations (yellow) are used for Si and SLG: Ti(20nm)/Pt(20nm)/Au(30nm) on Si, Pd(50nm) on SLG}
\label{fig:fig1}
\end{figure}

Here we demonstrate a GPM included in a Mach-Zehnder interferometer (MZI),Fig.\ref{fig:fig1}a, in which PM is realized in the form of a Si-insulator-SLG (SISLG) capacitor and the Si waveguide is used as a gating electrode to the SLG, Fig.\ref{fig:fig1}b). By applying a bias to the capacitor, charge is accumulated on the SLG electrode, shifting its Fermi level, E$_F$, and modifying its complex conductivity\cite{Wang2008,Falkovski2008,Stauber2008,Sorianello2016}. In this way, both the effective index and the optical loss of a SLG-functionalized Si waveguide can be tuned\cite{Sorianello2016}. If the SLG is doped beyond the Pauli blocking condition (E$_F>$0.4eV at 1.55$\mu$m), the device operates in the low-loss "transparency" region, where the only loss contribution is given by the Si waveguide propagation loss without SLG. In this regime, PM is dominant with respect to amplitude changes\cite{Sorianello2015}, with an enhanced modulation efficiency (V$_{\pi}$L$<$0.26Vcm)\cite{Sorianello2015} compared to typical Si photonics modulators operating in depletion mode (V$_{\pi}$L$>$2Vcm)\cite{Reed2008}.
\begin{figure}
\centerline{\includegraphics[width=90mm]{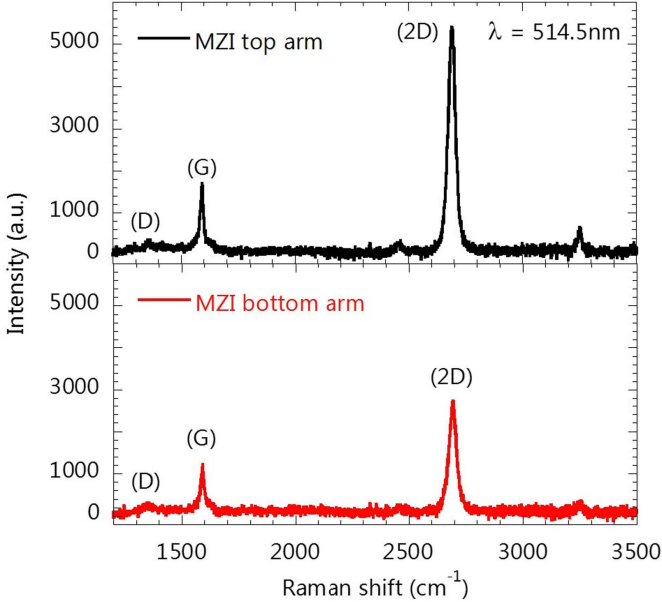}}
\caption{Raman spectra of SLG on the top (black curve) and bottom (red curve) arms of our GPM}
\label{fig:fig2}
\end{figure}

Our device is fabricated in a standard Si photonics platform using a Si-on-insulator (SOI) substrate (see Methods). The photonic structure consists of a balanced MZI with 3dB multimode interference (MMI) couplers\cite{Soldano1995} to split and combine the optical beam, and grating couplers\cite{Roelkens2006} to couple light in and out of the device, Fig.\ref{fig:fig1}a). The Si waveguide is designed to support a single transverse electric (TE) in-plane polarized optical mode. The waveguide is B doped to reduce the Si resistance and achieve high-speed operation\cite{Reed2008}. SLG is grown by chemical vapor deposition (CVD) and then wet transferred (see Methods) onto 10nm planarized Si dioxide (SiO$_2$) top cladding, Fig.\ref{fig:fig1}b). We use different SLG lengths onto the two MZI arms (300 and 400$\mu$m) to intentionally introduce a bias phase difference in the balanced interferometer structure for characterization purposes. The quality and uniformity of SLG after device fabrication are characterized by Raman spectroscopy (see Methods). The spectra show negligible D to G intensity ratio (Fig.\ref{fig:fig2}), indicating that no significant degradation and/or defects are introduced during the fabrication process\cite{Cancado2011,Bruna2014}. From the Raman G peak position, Pos(G), full width at half maximum, FWHM(G), and intensity and area ratios of D and G peaks we estimate\cite{Basko2009, Das2008} E$_F<$100meV ($\sim$200meV) at the longer top (shorter bottom) arms of the MZI.
\begin{figure}
\centerline{\includegraphics[width=75mm]{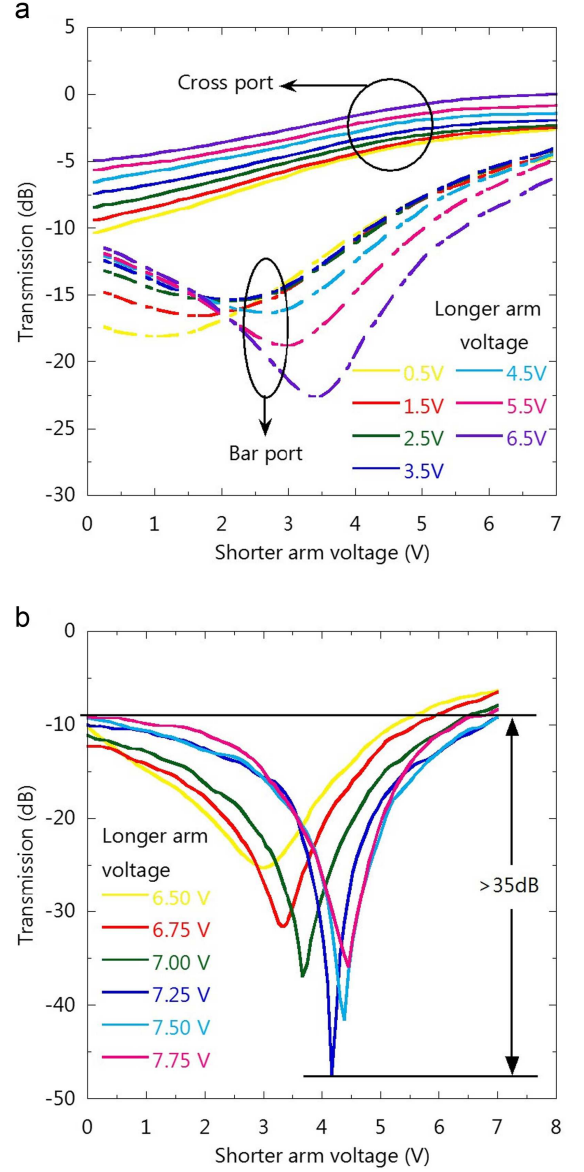}}
\caption{a) Light output at 1550nm as measured with a power meter at the bar port (solid line) and cross port (dashed lines) as a function of DC voltage applied to the shorter GPM (bottom arm in Fig.1a). Different colors refer to different DC voltages applied to the longer GPM (top arm in Fig.\ref{fig:fig1}a). b) 35dB extinction ratio at the bar port at 1550nm. By applying a 4.1V DC bias to the shorter GPM and 7.25V DC bias to the longer one, the phase difference between the two MZI arms approaches $\pi$ and the extinction is maximized.}
\label{fig:fig3}
\end{figure}

To test the electro-refractive effect, we first perform static characterizations by measuring the MZI output power as a function of bias applied to the GPMs on both MZI arms, Fig.\ref{fig:fig3}. A 1550nm laser source is coupled to the MZI input using a SMF, while the optical power at the MZI outputs, i.e the bar and cross ports, is collected with a similar fibre and monitored by a power meter (see Methods). In our balanced MZI the cross port is expected to be at maximum when the phase difference between the two arms is zero, while the bar port tends to be at zero power\cite{Born1970}. We measure an output power difference$\sim$5dB between the two ports when no voltage is applied. We assign this imbalance to the difference in absorption and phase accumulation between the two MZI arms caused by the different SLG lengths, doping and defects, Fig.\ref{fig:fig2}. By applying a bias to the SISLG capacitors, we decrease the bar port power to zero, with an extinction ratio (ER, ratio between maximum and minimum of the transmission)$>$35dB, Fig.\ref{fig:fig3}b). This is due to the phase change introduced in the two arms by electrical gating the SLGs. The 35dB ER is due to a phase difference between the two arms approaching $\pi$, and this is evidence of interferometric behavior\cite{Born1970}, with a considerable electro-refractive effect. In particular, we measure V$_{\pi}\sim$7.25V on the 400$\mu$m SISLG capacitor, corresponding to V$_{\pi}$L$\sim$0.28, very close to the theoretical prediction\cite{Sorianello2015}. This is at least a 8-fold improvement compared to state-of-the-art p-n junction based Si MZI modulators\cite{Denoyer2015}. The V$_{\pi}$L efficiency can be doubled by using a SLG-insulator-SLG phase modulator taking advantage of the effect of two SLGs on the effective index of the waveguide\cite{Sorianello2015}.
\begin{figure}
\centerline{\includegraphics[width=80mm]{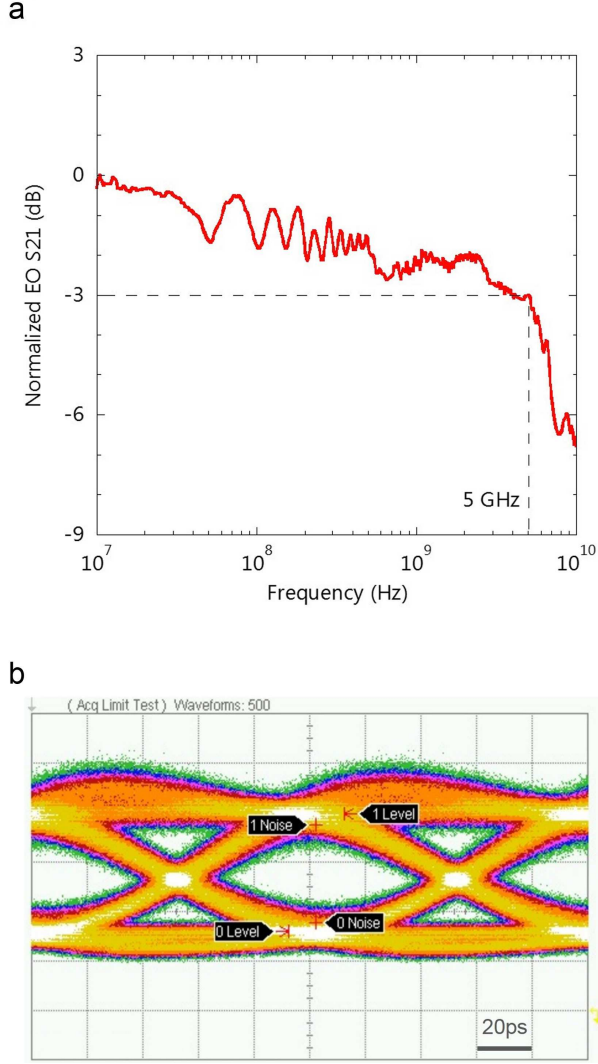}}
\caption{a) Electro-optical S21 bandwidth of our GMP. This is limited to 5GHz by the large contact resistance of the metal SLG contact. b) 2$^{31}$-1 PRBS NRZ eye diagram at 10Gb/s. The time scale is 20ps. The wavelength is 1550nm. The two GPMs are biased at 3.5V DC and driven with a 2V peak to peak RF signal. The eye diagram exhibits 3.94dB ER and 5.85dB SNR}
\label{fig:fig4}
\end{figure}

To test the modulator electro-optical (EO) bandwidth and operation speed, we first characterize the frequency response by using an electrical vector network analyzer (VNA) (see Methods). We measure a 3dB roll-off frequency of 5GHz at 4V, Fig.\ref{fig:fig4}a). The bandwidth is limited by the RC time constant primarily due to the SISLG capacitor and the series resistances. The overall capacitance C is the sum the capacitor on top of the waveguide core (C$_{ox}=\epsilon_0 \epsilon_{SiO_2} w_{wg}/t_{ox}\sim1.6$fF/$\mu$m, where $\epsilon_0$ is the vacuum dielectric constant, $\epsilon_{SiO_2}$ is the SiO$_2$ dielectric constant, w$_{wg}$ is the waveguide core width and t$_{ox}$ the oxide thickness) and the parasitic capacitor C$_p$ due to the SLG overlap on the Si slab (C$_p=\epsilon_0\epsilon_{SiO_2} w_{ol}/t_p \sim0.2$fF$\mu$m, where w$_{ol}\sim1\mu$m is the overlap and t$_p$ the oxide thickness between SLG and Si slab). C$_p$ depends on the alignment between transferred SLG and waveguide core, and it can contribute up to 20\% of the overall C. The series resistance, R, stems from different contributions: the Si resistance (R$_{si}$, estimated$\sim$5k$\Omega \cdot \mu$m, see Methods), the SLG lead from the waveguide core to the metal contact, R$_g$ (estimated$<$1k$\Omega \cdot \mu$m, see Methods), the metal to SLG contact resistance, R$_c$. The latter is estimated as$\sim$10k$\Omega \cdot \mu$m by using a transfer length measurement (TLM) on test samples (see Methods). In our device, the metal/SLG contact resistance is the main factor limiting the radio-frequency (RF) bandwidth. This can be improved by reducing C$_p$ and R$_c$. Contact resistances down to 100$\Omega \cdot \mu$m have been reported\cite{Leong2014}, such values would increase the RF bandwidth up to 30GHz and beyond, leading to high frequency operations comparable with state-of-the-art modulators\cite{Xiao2013, Streshinsky2013, Denoyer2015,Reed2013, Dong2012,Fresi2016,Milivojevic2013}.
\begin{figure}
\centerline{\includegraphics[width=80mm]{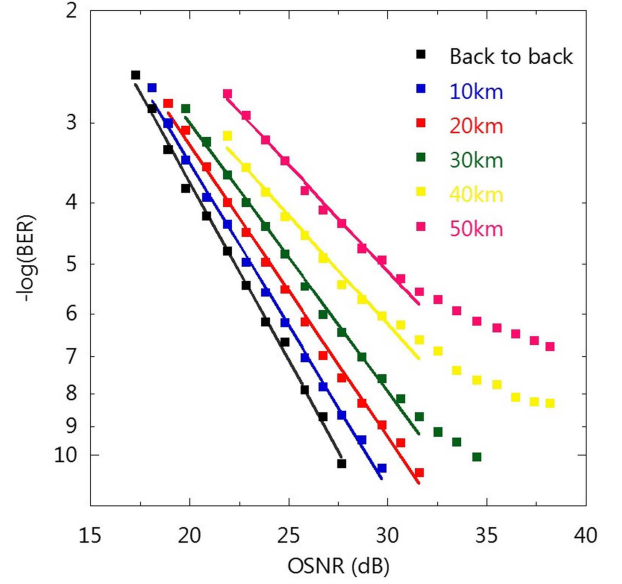}}
\caption{BER measurements of the 2$^{31}$-1 PRBS NRZ signal at 10Gb/s (1550nm wavelength) as a function of OSNR in back-to-back and after transmission over different standard SMF spool lengths. BER below 10$^{-10}$ is achievable up to 30km. Considering state-of-the-art systems employing SD-FEC, enabling pre-FEC BER threshold in the range of 10$^{-3}$, the device demonstrates error free operation up to 50km SMF transmission}
\label{fig:fig5}
\end{figure}

We now discuss the optical response when the MZI modulator is driven by a non-return-to-zero (NRZ) signal\cite{Seimetz2009}. We measure the MZI NRZ eye diagram with the two GPMs biased at 3.5V and operating in push-pull configuration, with 2V peak-to-peak and 2$^{31}$-1 pseudorandom binary sequence (PRBS)\cite{Agrawal2010} signals. Fig.\ref{fig:fig3}b plots an open eye diagram at 10Gb/s with$\sim$4dB ER and$\sim$6dB signal-to-noise ratio (SNR). With a random bit stream, the individual GPM average energy consumption is given by\cite{Miller2012} C($\Delta$V)$^2$/4, where $\Delta$V is the voltage variation driven to the contacts in order to charge/discharge the GPM capacitance. In our device, we get$\sim$1pJ/bit. The energy consumption could be reduced with the SLG-insulator-SLG configuration, potentially allowing us to halve the driving voltage\cite{Sorianello2015}.

We then carry out bit error rate (BER) measurements when the NRZ signal is transmitted in a standard single mode fiber (SMF) over different distances, Fig.\ref{fig:fig5}. We measure the BER as a function of optical signal-to-noise ratio (OSNR) in back-to-back configuration, and after propagation in SMF spools of different lengths from 10 to 50km. We have error free operation (BER$<$10$^{-10}$) up to 30km, while at longer propagation distances a BER floor appears. However, considering the state-of-the-art systems currently employing a soft-decision forward error correction (SD-FEC)\cite{Infinera2015}, with overhead in the 7-25\% range, and enabling a pre-FEC BER threshold up to 3.4x10$^{-2}$ or even higher\cite{Rahman2015}, our modulator exhibits error free operation up to 50km.

In summary, we demonstrated a graphene-silicon phase modulator operating in the GHz regime. We included the GPM in a MZI device demonstrating a static modulation depth of 35dB and modulation efficiency of 0.28Vcm, outperforming state-of-the-art Si-based p-n junctions and comparable to the SIS-capacitor based Si modulators. The modulator operates at 10Gbit/s, showing an open eye diagram and error free transmission over 50km single mode fibre. These results pave the way to the realization of graphene modulators for a wide range of telecom applications where the phase modulation is crucial.

We acknowledge funding from the Graphene Flagship Project ID: 696656, ERC Grant Hetero2D and EPSRC Grants EP/ 509 K01711X/1, EP/K017144/1, EP/N010345/1, EP/M507799/ 5101 and EP/L016087/1. We acknowledge Graphenea for the provision of CVD samples.
\section{Methods}
The Si photonic device is prepared on the IMEC iSiPP25G Si on insulator (SOI) platform\cite{Absyl2015}. The MZI interferometer is based on Si ridge waveguides with 60nm slab and a core cross-section 480nm$\times$220nm. SLG is placed on the Si waveguide separated by a 10nm spacer of high quality thermal SiO$_2$. Graphene is grown by CVD on Cu foils, as described in Ref.\cite{Li2009}, and then transferred onto our Si samples following the procedure described in Ref.\cite{Zurutuza2015}. Graphene patterning is done by using a bilayer stack of PMMA/IX845, while etching is done using conventional O$_2$-plasmas. The metal contacts on SLG and Si are processed separately in two consecutive steps. SLG is contacted with a single 50nm thick Palladium (Pd) layer, while the metal contact on Si consists of a Titanium (Ti)/Platinum (Pt)/Gold (Au) stack with thickness 20nm/20nm/30nm deposited on the Si surface after HF cleaning to remove the native SiO$_2$. Contact resistance and mobility are evaluated on different samples with equivalent processing. Typically values of 10k$\Omega\cdot\mu$m are extracted at the charge neutrality point using the TLM method\cite{Politou2016,Politou2017}, with mobility$\sim$1500cm$^2$V$^{-1}$s$^{-1}$.

After device fabrication, the quality and uniformity of SLG at both arms of the MZI is monitored by Raman spectroscopy using a Horiba LabRam Evolution Raman spectrometer with 514.5nm laser and optical power below 0.1mW. The Raman spectra from the top and bottom arms (GPMs) of the MZI are normalized to the Si Raman peak at 521cm$^{-1}$. The 2D peak is a single sharp Lorentzian with full width at half maximum FWHM(2D)$\sim$36cm$^{-1}$ ($\sim$39cm$^{-1}$) at the top (bottom) arm of the MZI, a signature of SLG\cite{Ferrari2013}. The spectra show negligible I(D)/I(G)$\sim$0.12 ($\sim$0.16), indicating that no significant degradation and/or defects are introduced during the fabrication process\cite{Cancado2011}. We estimate a defects density$\sim$2.3$\cdot$10$^{10}$cm$^{-2}$ and$\sim$6.5$\cdot$10$^{10}$cm$^{-2}$ at top and bottom GPMs respectively\cite{Cancado2011,Bruna2014}. Pos(G) is$\sim$1592~cm$^{-1}$ ($\sim$1593cm$^{-1}$), with FWHM(G)$\sim$17cm$^{-1}$ ($\sim$19cm$^{-1}$). The 2D peak position, Pos(2D) is$\sim$2691cm$^{-1}$ ($\sim$2693cm$^{-1}$), while the 2D to G peak intensity and area ratios I(2D)/I(G) and A(2D)/A(G), are$\sim$3.7 (2.6) and 8.2 (5.3), indicating a p-doping$<$100meV ($\sim$200meV)\cite{Basko2009,Das2008} at the top (bottom) arm of the MZI. These correspond to carriers concentrations$<$1$\cdot$10$^{12}$cm$^{-2}$ ($\sim$2.3$\cdot10^{12}$~cm$^{-2}$)\cite{Das2008}.

R$_{si}$ is evaluated by numerical electrical simulations of the GPMs cross section with a commercial-grade device simulator that self-consistently solves the Poisson and drift-diffusion equations\cite{lumerical}. We use Si doping and resistivity as obtained from the IMEC iSiPP25G technology\cite{Absyl2015}. For SLG, we estimate a sheet resistance $<$400$\Omega/\square$ when the SLG is gated beyond the Pauli blocking condition (E$_F>$0.4eV). For E$_F>$0.4eV the carrier concentration exceeds 1.3$\cdot$10$^{13}$cm$^{-2}$ ($\left|n_s\right|=\pi^{-1}\cdot (E_F/(\hbar v_F))^2$, where n$_s$ is the carrier concentration, $\hbar$ the bar Planck constant and v$_F$ is the Fermi velocity\cite{Wang2008}), the mobility degradation from the ungated value is negligible\cite{Hirai2014}. As the SLG lead from the waveguide core to the metal contact is 2.5$\mu$m, R$_g$ can be estimated$<$400$\Omega/\square \cdot 2.5\mu$m=1k$\Omega \cdot \mu$m.

The input/output optical coupling is obtained through cut single mode fibres with cleaved output surfaces. The input fibre is connected to the laser source with a fibre polarization controller to maximize the input coupled light. We use a tunable external cavity laser fixed at 1550nm. The output fibre is connected to a high sensitivity power meter to measure the static characteristics in Fig.\ref{fig:fig3}. Two ground signal (GS) high frequency probes apply DC and RF signals to the two MZI GPMs. For small-signal RF bandwidth measurements we use an electrical VNA. Its output is connected through a 50$\Omega$ matched high frequency cable to the GS probe contacting the GPM. We use a bias-tee to combine RF power and DC bias. We set a DC bias of 4V and an RF power of -17dBm. The light at the output of the MZI is modulated by the RF signal from the VNA and collected by a low noise, high frequency photodetector connected to the VNA input. The signals for the eye diagram and BER measurements are generated by a pattern generator (PG) and collected with a digital sampling oscilloscope (eye diagram) and a BER tester. The PG provides the 2$^{31}$-1 pseudorandom binary sequence (PRBS) at 10Gb/s. The signal and inverted signal are sent to the two GPMs through the RF cable and bias-tee. We use the optical input of the oscilloscope to collect the light out of the MZI GPM and visualize the eye diagram of Fig.\ref{fig:fig4}b. For BER evaluation, we used a high frequency photo-receiver collecting the light at the output of the modulator and connected to the BER tester electrical input.

\end{document}